\def\Draft{}
\newcommand{\AuthorA}{Laurent U.~Perrinet}%
\newcommand{\AuthorB}{Guillaume S.~Masson}%
\newcommand{\Address}{Institut de Neurosciences de la Timone, UMR 7289, CNRS / Aix-Marseille University - 27, Bd. Jean Moulin, 13385 Marseille Cedex 5, France}%
\newcommand{\Website}{http://invibe.net/LaurentPerrinet}%
\newcommand{\Title}{Motion-based prediction is sufficient to solve the aperture problem}%
\newcommand{\Abstract}{%
In low-level sensory systems, it is still unclear how the noisy information collected locally by neurons may give rise to a coherent global percept. This is well demonstrated for the detection of motion in the aperture problem: as luminance of an elongated line is symmetrical along its axis, tangential velocity is ambiguous when measured locally. Here, we develop the hypothesis that motion-based predictive coding is sufficient to infer global motion. Our implementation is based on a context-dependent diffusion of a probabilistic representation of motion. We observe in simulations a progressive solution to the aperture problem similar to physiology and behavior. We demonstrate that this solution is the result of two underlying mechanisms. First, we demonstrate the formation of a tracking behavior favoring temporally coherent features independently of their texture. Second, we observe that incoherent features are explained away while coherent information diffuses progressively to the global scale. Most previous models included ad-hoc mechanisms such as end-stopped cells or a selection layer to track specific luminance-based features as necessary conditions to solve the aperture problem. Here, we have proved that motion-based predictive coding, as it is implemented in this functional model, is sufficient to solve the aperture problem. This solution may give insights in the role of prediction underlying a large class of sensory computations. %
}%
\newcommand{\Keywords}{motion detection; aperture problem; probabilistic representation; predictive coding; association field}%
\newcommand{\Acknowledgments}{
This work is supported by EC IP project FP6-015879, ``FACETS'' and FP7-269921, ``BrainScaleS''. %
Code to reproduce figures and supplementary material are available on the corresponding author's website at \url{\Website/Publications/Perrinet12pred} %
}%
\documentclass[pdftex,a4paper,\Draft]{article}
\usepackage{microtype}

\newcommand{\head}[1]{}

\usepackage[comma,sort&compress,square]{natbib}
\usepackage{units}%
\usepackage{amsmath}
\usepackage{graphicx,fltpage}%
\usepackage[pdfusetitle,colorlinks=false,pdfborder={0 0 0}]{hyperref}%
\title{\Title}%
\author{\AuthorA\ and \AuthorB \thanks{\Address}}%
\date{}%

\begin{document}%
\maketitle%
\ifthenelse{\equal{\Draft}{draft}}{\vspace{-1cm}\tableofcontents \newpage }{}
\begin{abstract}
\Abstract
\end{abstract}%
\section*{Keywords}
\Keywords
\section{Introduction}
\subsection{Problem statement}
A central challenge in neuroscience is to explain how local information that is represented in the activity of single neurons, can be integrated to enable global and coherent responses at population and behavioral levels. A classical illustration of this problem is given by the early stages of visual motion processing. Visual cortical areas, such as the primary visual cortex (V1) or the medio-temporal (MT) extra-striate area can extract geometrical structures from luminance changes that are sensed by large populations of direction- and speed-selective neurons within topographically organized maps~\citep{Hildreth87}. However, these cells have only access to the limited portion of the visual space falling inside their classical receptive fields. By consequence, local information is often incomplete and ambiguous, as for instance when measuring the motion of a long line that crosses their receptive field. Because of the symmetry along the line's axis, the measure of the tangential component of translation velocity is completely ambiguous, leading to the \emph{aperture problem} (see Figure~\ref{fig:one}-A). As a consequence, most V1 and MT neurons indicate the slowest element from the family of vectors compatible with the line's translation, that is the speed perpendicular to the line orientation~\citep{Albright84}. These neurons are often called component-selective cells and can signal only orthogonal motions of local 1D edges from more complex moving patterns. Integrated in area MT, such local preferences introduce biases in the estimated direction and speed of the translating line. A behavioral consequence is that perceived direction of an elongated tilted line is initially biased towards the motion direction orthogonal to its orientation~\citep{Lorenceau92,Masson02a,Wallace05,Born06,Pei10}. There are however other MT neurons, called pattern selective cells, that can signal the true translation vector corresponding to such complex visual patterns and, hence drive correct, steady-state behaviors~\citep{Movshon85,Rodman89,Pack01}. Ultimately, these neurons provide a solution similar to the interception-of-constraints (IOC)~\citep{Fennema79,Adelson82} by combining the information of multiple component cells (for a recent model, see~\citep{Bowns11}). %

The classical view is that these pattern selective neurons integrate information from a large pool of component cells signaling a wide range of directions, spatial frequencies, speeds and so on~\citep{Rust06}. However, this two stage, feed-forward model of motion integration is challenged by several recent studies that call for more complex computational mechanisms (see~\citep{Masson10} for reviews). First, there are neurons outside area MT that can solve the aperture problem. For instance, V1 end-stopped cells are sensitive to particular features such as line-endings and can therefore signal unambiguous motion at a much smaller spatial scale providing that the edge falls within their receptive field~\citep{Pack04}. These neurons could contribute to pattern selectivity in area MT~\citep{Tsui10} but this solution only pushes the problem back to earlier stages of cortical motion processing since one must now explain the emergence of end-stopping cells. Second, all neural solutions to the aperture problem are highly dynamical and build up over dozens of milliseconds after stimulus onset~\citep{Pack01,Pack03,Pack04,Smith10}. This can explain why perceived direction of motion gradually changes over time, shifting from component to pattern translation~\citep{Lorenceau92,Masson02a,Wallace05}. Classical feedforward models cannot account for such temporal dynamics and its dependency upon several properties of the input such as contrast or bar length~\citep{Rust06,Tsui10}. Third, classical computational solutions ignore the fact that any object moving in the visual world at natural speeds will travel across many receptive fields within the retinotopic map. Thus, any single, local receptive field will be stimulated over a period of time that is much less that the time constants reported above for solving the aperture problem (see~\citep{MassonMontagnini10}). Still, single neuron solutions for ambiguous motion that have been documented so far only with conditions where the entire stimulus is presented within the receptive field~\citep{Pack04} and with the same geometry~\citep{Majaj07} over dozens of milliseconds. 

Thus, there is an urgent need for more generic computational solutions. We have recently proposed that diffusion mechanisms within a cortical map can solve the aperture problem without the need for complex local mechanisms such as end-stopping or pooling across spatio-temporal frequencies~\citep{Tlapale10vr,Tlapale11}. This approach is consistent with the role of recurrent connectivity in motion integration~\citep{Bayerl04} and can simulate the temporal dynamics of motion integration in many different conditions. Moreover, it can reverse the perspective that is dominant in feedforward models where local properties such as end-stopping, pattern selectivity or other types of extra-classical receptive fields phenomena are implemented by built-in, specific neuronal detectors. Instead, these properties can be seen as solutions emerging from the neuronal dynamics of the intricate, recursive contributions of feed-forward, feedback and lateral interactions. A vast theoretical, and experimental challenge is therefore to elucidate how diffusion models can be implemented by realistic populations of neurons dealing with noisy inputs.

The aperture problem in vision must be seen as an instance of the more generic problem of information integration in sensory systems. The aperture problem, as well as the correspondence problem, can be seen as a class of under-constrained inverse problems faced by many different sensory and cognitive systems. Interestingly, recent experimental evidence have pointed out strong similarities in the dynamics of the neural solution for spatio-temporal integration of information in space. For instance, there is a tactile counterpart of the visual aperture problem and neurons in the monkey somatosensory cortex exhibit similar temporal dynamics to that of area MT neurons~\citep{Pei08,Pei10,Pei11}. These recent results urge the need to build a theoretical framework that can unify these generic mechanisms such as robust detection and context-dependent integration and to propose a solution that would apply to different sensory systems. An obvious candidate is to build \emph{association fields} that would gather neighboring information such as to enhance constraints on the response. This is the goal of our study to provide a theoretical framework using probabilistic inference. %

\begin{FPfigure}%
\centering{\includegraphics[width=\textwidth]{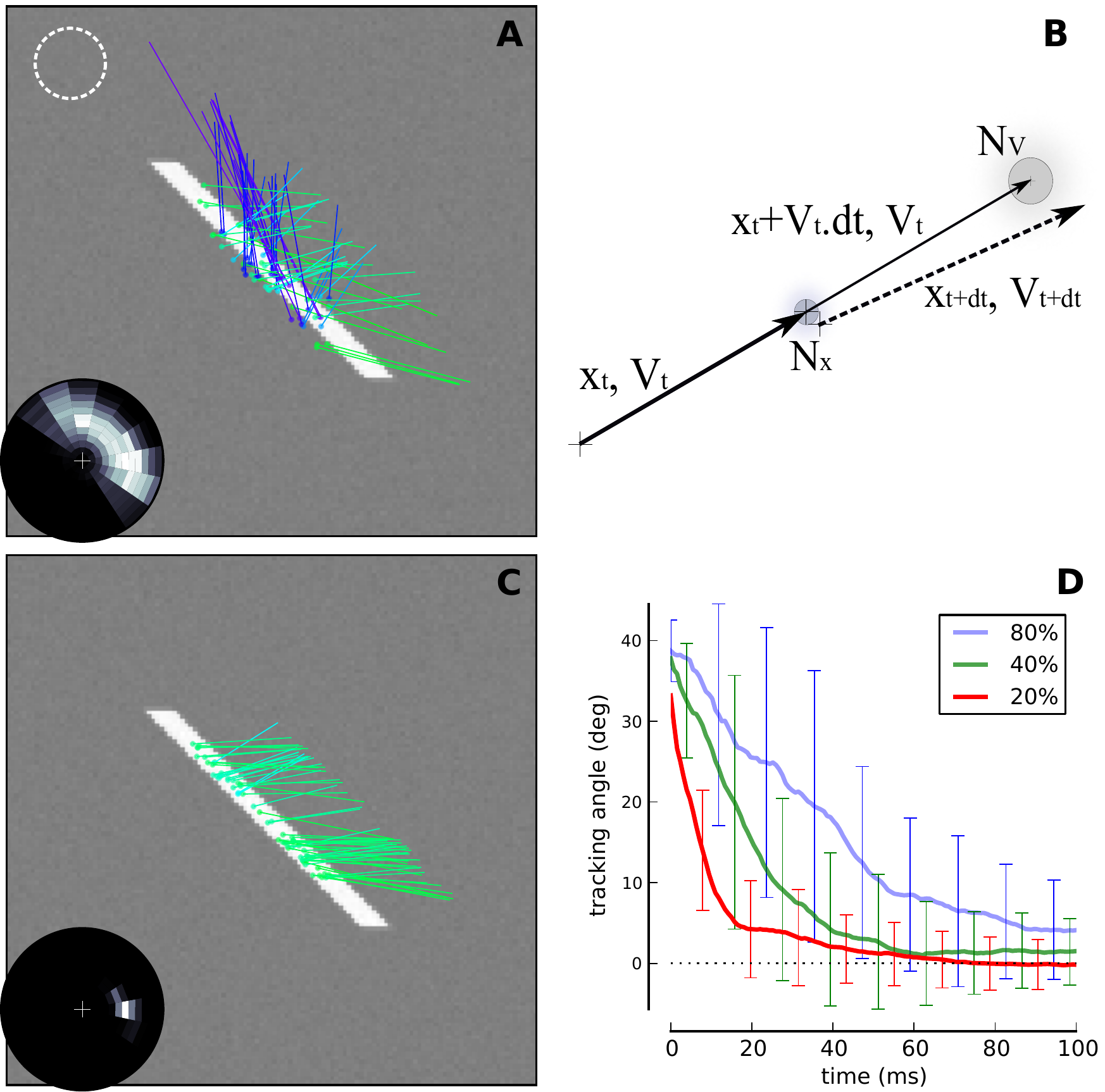}}%
\caption{%
\textbf{(A)} The estimation of the motion of an elongated, slanted segment (here moving horizontally to the right) on a limited area (such as the dotted circle) leads to ambiguous velocity measurements compared to physical motion: it's the \emph{aperture problem}. We represent as arrows the velocity vectors that are most likely detected by a motion energy model; hue indicates direction angle. Due to the limited size of receptive fields in sensory cortical areas (such as shown by the dotted white circle), such problem is faced by local populations of neurons that visually estimate the motion of objects. \textbf{(A-inset)} On a polar representation of possible velocity vectors (the cross in the center corresponds to the null velocity, the outer circle corresponding to twice the amplitude of physical speed), we plot the empirical histogram of detected velocity vectors. This representation gives a quantification of the aperture problem in the velocity domain: At the onset of motion detection, information is concentrated along an elongated constraint line (white=high probability, black=zero probability). \textbf{(B)} We use the prior knowledge that in natural scenes, motion as defined by its position and velocity is following smooth trajectories. Quantitatively, it means that velocity is approximately conserved and that position is transported according to the known velocity. We show here such a transition on position and velocity (respectively $\vec{x}_{t}$ and $\vec{V}_{t}$) from time $t$ to $t + dt$ with the perturbation modeling the smoothness of prediction in position and velocity (respectively $\mathcal{N}_x$ and $\mathcal{N}_V$). \textbf{(C)} Applying such a prior on a dynamical system detecting motion, we show that motion converges to the physical motion after approximately one spatial period (the line moved by twice its height). \textbf{(C-Inset)} The read-out of the system converged to the physical motion: Motion-based prediction is sufficient to resolve the aperture problem. \textbf{(D)} As observed at the perceptual level~\citep{Castet93a,Pei10}, size and duration of the tracking angle bias decreased with respect to the height of the line. Height was measured relative to a spatial period (respectively 60\%, 40\% and 20\%). Here we show the average tracking angle red-out from the probabilistic representation as a function of time, averaged over 20 trials (error bars show one standard deviation). %
\label{fig:one}%
}%
\end{FPfigure}%

Herein, we explore the hypothesis that the aperture problem can be solved thanks to predictive coding. We introduce a generic probabilistic framework for motion-based prediction as a specific dynamical spatio-temporal diffusion process on motion representation, as originally proposed by~\citet{Burgi00}. However, we do not perform an approximation of the dynamics of probabilistic distributions using a neural network implementation, as they did. Instead, we develop a method to simulate precise predictions in topographic maps. We test our model against behavioral and neuronal results that are signatures of the key properties of primate visual motion detection and integration. Furthermore, we demonstrate that several properties of low-level motion processing (i.e. feature motion tracking, texture-independent motion, context-dependent motion integration) naturally emerge from predictive coding within a retinotopic map. Lastly, we discuss the putative role of prediction in generic neural computations.%
\subsection{Probabilistic detection of motion}%
First, we define a generic probabilistic framework for studying the aperture problem and its solution. Translation of an object in the planar visual space at a given time is fully given by the probability distribution of its position and velocity, that is, as a distribution of our value of belief among a set of possible velocities. It is usual to define motion probability at any given location. If one particular velocity is certain, its probability becomes $1$ while other probabilities are $0$. The more the measurement is uncertain (for instance when increasing noise), the more the distribution of probabilities will be spread around this peak. This type of representation can be successfully used to solve a large range of problems related to visual motion detection. These problems belong in all generality to optimal detection problems of a signal perturbed by different sources of noise and ambiguity. In particular, the aperture problem is explicitly described by an elongated probability distribution function (PDF) along the constraint defined by the orientation of the line (see Figure~\ref{fig:one}-A, inset). This constitutes an ill-posed inverse problem as different possible velocities may correspond to the physical motion of the line. %

In such a framework, Bayesian models make explicit the optimal integration of sensory information with prior information. These models may be decomposed in three stages. First, one defines likelihoods as a measure of belief knowing the sensory data. This likelihood is based on the definition of a generative model. Second, any prior distribution, that is, any information on the data that is known before observing it, may be combined to the likelihood distribution to compute a posterior probability using Bayes' rule. The prior defines generic knowledge on the generative model over a set of inputs, such as regularities observed in the statistics of natural images or behaviorally relevant motions. Finally, a decision can be made by optimizing a behavioral cost dependent on this posterior probability. An often used choice is to choose the belief that corresponds to the maximum a posteriori probability. The advantage of Bayesian inference compared to other heuristics is that it explicitly states qualitatively and quantitatively all hypotheses (generative models of observation noise and of the prior) that lead to a solution. %
\subsection{Luminance-based detection of motion}%
Such a Bayesian scheme can be applied to motion detection using a generative model of the luminance profile in the image. This is first based on the luminance conservation equation. Knowing the velocity $\vec{V}$, we can assume that luminance is {approximately} conserved along this direction, that is, that after a small lapse $dt$: 
\begin{equation}%
I_{t + dt}(\vec{x} + \vec{V} \cdot dt) = I_t(\vec{x}) + \mathcal{N}_I \label{eq:obs}%
\end{equation}
where we define luminance at time $t$ by $I_t(\vec{x})$ as a function of position $\vec{x}$ and $\mathcal{N}_I$ is the observation noise. Using the Laplacian approximation, one can derive the likelihood probability distribution $p(I_t(\vec{x}) | \vec{V})$ as a Gaussian distribution. In such a representation, precision is finer for a lower variance. Indeed, it is easy to show that the logarithm of $p(I_t(\vec{x}) | \vec{V})$ is proportional to the output of a correlation-based elementary motion sensors or equivalently to a motion-energy detector~\citep{Adelson85}. Second,~\citet{Weiss02} showed that using a prior distribution $p(\vec{V})$ that favors slow speeds, one could explain why the initial perceived direction in the aperture problem is perpendicular to the line. Interestingly, lower contrast motion results in wider distributions of likelihood and thus posterior $p(\vec{V} | I_t(\vec{x}) )$. Therefore, contrast dynamics for a wide variety of simple motion stimuli is determined by the shape of the probability distribution (i.e. Gaussian-like distributions) and the ratio between variances of likelihood and prior distributions as was validated experimentally on behavioral data~\citep{Barthelemy08}. With ambiguous inputs, this scheme gives a measure consistent with our formulation of the aperture problem, where probability is distributed along a constraint line defined by the orientation of the line (see Figure~\ref{fig:one}-A, inset).%

\head{\emph{spatial integration:} local geometry / independence / neighbors dependence}
The generative model explicitly assumes a translational motion $\vec{V}$ over the observation aperture, such as the receptive field of a motion-sensitive cell. Usually, a distributed set $\vec{V}_{t}(\vec{x})$ of motion estimations at time $t$ over fixed positions $\vec{x}$ in the visual field gives a fair approximation of a generic, complex motion that can be represented in a retinotopic map such as V1/MT areas. This provides a field of probabilistic motion measures $p(I_t(\vec{x}) | \vec{V}_{t}(\vec{x})))$. To generate a global read-out from these local informations, we may integrate these local probabilities over the whole visual field. Assuming independence of the local information as in~\citet{Weiss02}, spatio-temporal integration is modeled at time $T$ by Equation~\eqref{eq:obs} and $p(\vec{V} | I_{0:T} ) \propto \prod_{\vec{x}, 0 \leq t \leq T} p(I_t(\vec{x}) | \vec{V}(\vec{x})) p(\vec{V}) $, where we write as $I_{0:t}$ the information on luminance from time $0$ to $t$. Such models of spatio-temporal integration can account for several nonlinear properties of motion integration such as monotonic spatial summation and contrast gain control and are successful in explaining a wide range of neurophysiological and behavioral data. In particular, it is sufficient to explain the dynamics of the solution to the aperture problem if we assume that information from lines and line-endings was a priori segmented~\citep{Barthelemy08}. This type of model provides a solution similar to the vector average and we have previously shown that the hypothesis of an independent sampling cannot account for some non-linear aspects of motion integration such as super-saturation of the spatial summation functions, unless some \emph{ad hoc} mechanisms such as surround inhibition is added~\citep{Perrinet07neurocomp}. In the particular case of our definition of the aperture problem (see Figure~\ref{fig:one}-A), the information from such Bayesian measurement at every time step will always give the same probability distribution function (described by its mean $\vec{V}_m$ and variance $\Sigma$), where $\vec{V}_m$ shows a bias toward the perpendicular of the line (see Figure~\ref{fig:one}-A, inset). The independent integration of such information will therefore necessary lead to a  finer precision (the variance becomes $\Sigma/T$) but with always the same mean: The aperture problem is not solved. %
\subsection{Motion-based predictive coding}
Failure of the feedforward models in accounting for the dynamics of global motion integration originates from the underlying hypothesis of independence of motion signals in neighboring parts of visual space. The independence hypothesis set above formally states that the local measurement of global motion is the same everywhere, independently of the position of different motion parts. In fact, the independence hypothesis assumes that if local motion signals would be randomly shuffled in position, they would still yield the same global motion output (e.g.~\citep{Movshon85}). As shown by~\citet{Watamaniuk95}, this hypothesis is particularly at stake for motions along coherent trajectories: motion as a whole is more than the sum of its parts. A solution used in previous models solving the aperture problem is to add some additional heuristics, such as a selection process~\citep{Nowlan95,Weiss02} or a constraint that motion is relatively smooth away from luminance discontinuities~\citep{Tlapale10vr}. A first assumption is that the retinotopic position of motion is an essential piece of information to be represented. In particular, in order to achieve fine-grained predictions, it is essential to consider that spatial position of motion $\vec{x}$, instead of being a given parameter (classically, a value on a grid), is an additional random variable for representing motion along with  $\vec{V}$. Compared to the representation $p(\vec{V}(\vec{x}) | I )$ used in previous studies~\citep{Burgi00,Weiss02}, the probability distribution $p(\vec{x}, \vec{V} | I )$ more completely describes motion by explicitly representing its spatial position jointly with its velocity. Indeed, it is more generic as it is possible to represent any distribution $p(\vec{V}(\vec{x}) | I )$ with a distribution $p(\vec{x}, \vec{V} | I )$, while the reverse is not true without knowing the spatial distribution of the position of motion $p(\vec{x} | I )$. This introduces an explicit representation of the segmentation of motion in visual space which will be an essential ingredient in motion-based predictive coding. %

Here, we explore the hypothesis that we may take into account most dependence of local motion signals between neighboring times and positions by implementing a predictive dependence of successive measurements of motion along a smooth trajectory. In fact, we know \emph{a priori} that natural scenes are predictable due to both rigidity and inertia of physical objects. Due to the projection of their motion in visual space, visual objects preferentially follow smooth trajectories (see Figure~\ref{fig:one}-B). We may implement this constraint into a generative model by using the transport equation on motion itself. This assumes that at time $t$, during the small lapse $dt$, motion was translated proportionally to its velocity :%
\begin{align}%
\vec{x}_{t + dt} &= \vec{x}_{t} + \vec{V}_{t} \cdot dt + \mathcal{N}_{\vec{x}}\label{eq:dyn1}\\%
\vec{V}_{t + dt} &= \vec{V}_{t} + \mathcal{N}_{\vec{V}} \label{eq:dyn2}%
\end{align}%
where $\mathcal{N}_{\vec{x}}$ and $\mathcal{N}_{\vec{V}}$ are respectively position and velocity unbiased noises on the motion's trajectory. In the noiseless case, on the limit when $ dt$ tends to zero, this is the auto-advection term in the Navier-Stokes equations and thus implements a ``fluid'' prior in the inference of local motion. In fact, it is important to properly tune $\mathcal{N}_{\vec{x}}$ and $\mathcal{N}_{\vec{V}}$ since the variance of these distributions explicitly quantify the precision of the prediction (see Figure~\ref{fig:one}-B). 
\head{markov chain / nothing new: spatial kalman / balance diffusion-reaction }%

\begin{figure}%
\centering{\includegraphics[width=.99\columnwidth]{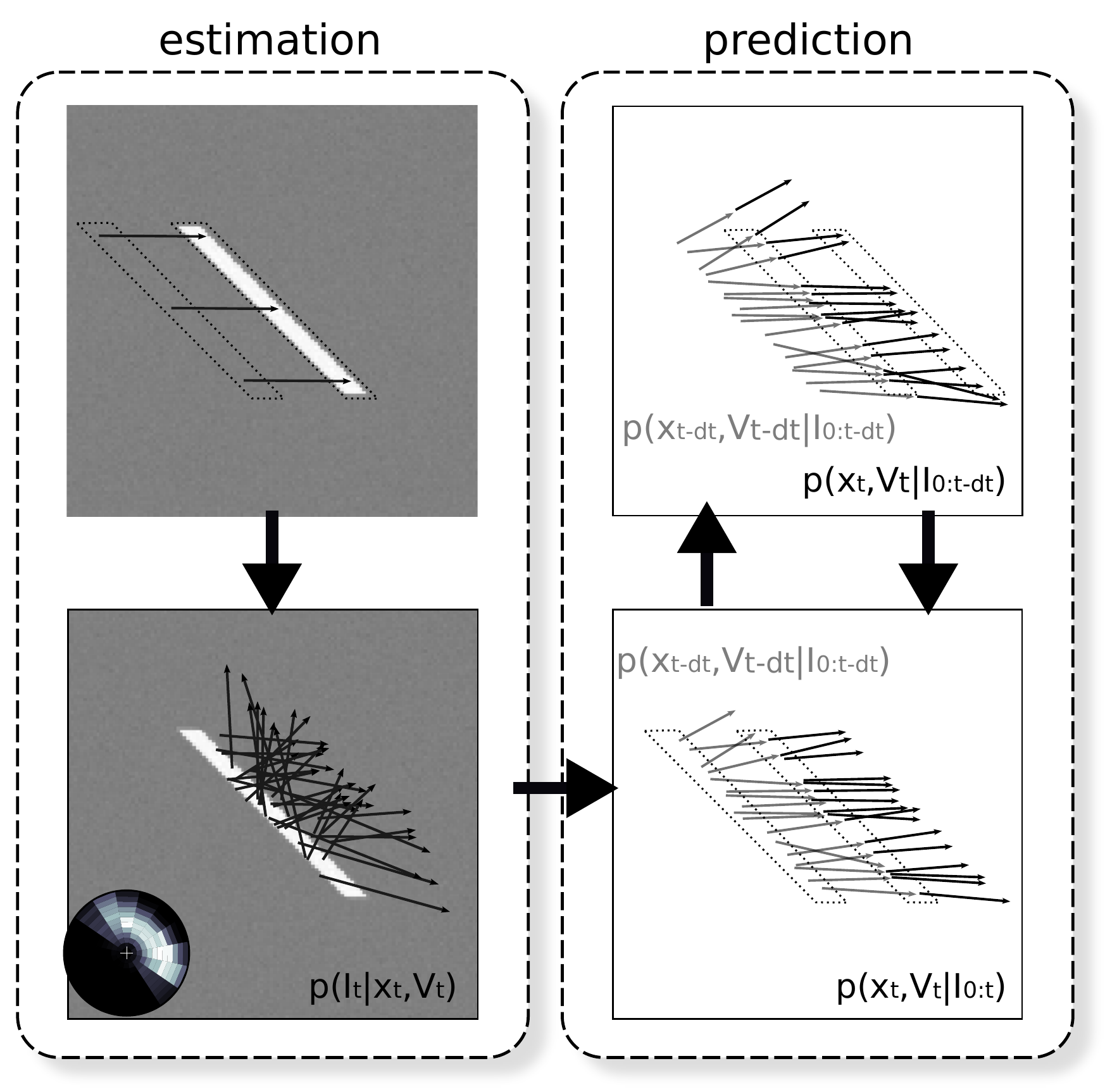}}%
\caption{Architecture of the model. %
The model is constituted by a classical measurement stage and of a predictive coding layer. The measurement stage consists of \textbf{(A)} inferring from two consecutive frames of the input flow, \textbf{(B)} a likelihood distribution of motion. This layer interacts with the predictive layer which consists of \textbf{(C)} a prediction stage that infers from the current estimate and the transition prior the upcoming state estimate and \textbf{(D)} an estimation stage that merges the current prediction of motion with the likelihood measured at the same instant in the previous layer (B).%
\label{fig:two}
}%
\end{figure}

We may now use this generative model to integrate motion information. Assuming for simplicity that sensory representation is acquired at discrete, regularly spaced times, let's define integration using a Markov random chain on joint random variables $z_{t}=\vec{x}_{t}, \vec{V}_{t}$: 
\begin{align}%
p(z_{t} | I_{0:t-dt}) & = \int_{dz_{t-dt}} p( z_{t} | z_{t-dt}) \cdot p(z_{t-dt} | I_{0:t-dt})  \label{eq:mrf1}\\%
p(z_{t} | I_{0:t}) & = p( I_{t} | z_{t}) \cdot p(z_{t} | I_{0:t-dt}) / p(I_{t} | I_{0:t-dt} ) \label{eq:mrf2}
\end{align}%
To implement this recursion, we first compute $p( I_{t} | z_{t})$ from the observation model (Equation~\eqref{eq:obs}). The predictive prior probability $p( z_{t} | z_{t-dt})$, that is, $p(\vec{x}_{t}, \vec{V}_{t} | \vec{x}_{t-dt}, \vec{V}_{t-dt})$ is defined by the generative model defined in Equation~\eqref{eq:dyn1} and \eqref{eq:dyn2}. Note that prediction (Equation~\eqref{eq:mrf1}) always increases the variance by ``diffusing'' information. On the other hand, during estimation (Equation~\eqref{eq:mrf2}), coherent data increases precision of the estimation while incoherent data increases the variance. This balance between diffusion and reaction will be the most important factor for the convergence of the dynamical system. Overall, these master equations, along with the definition of the prior transition $p( z_{t} | z_{t-dt})$, define our model as a dynamical system with a simple global architecture but yet with complex recurrent loops (see Figure~\ref{fig:two}). 

\head{ \textbf{BUT:} very hard to implement / as in Burgi association field :-) or... / SMC}%
Unfortunately, the dimensionality of the probabilistic representation makes impossible the implementation of realistic simulations of the full dynamical system on classical computer hardware. In fact, even with a moderate quantization of the relevant representation spaces, computing integrals over hidden variables in the filtering and prediction equations (respectively Equations~\eqref{eq:mrf1} and \eqref{eq:mrf2}) leads to a combinatorial explosion of parameters that is intractable with the limited memory of current sequential computers. Alternatively, if we assume that all probability distribution are Gaussian, this formulation is equivalent to Kalman filtering on joint variables. Such type of an implementation may be achieved using for instance a neuromorphic approximation of the above equations~\citep{Burgi00}. Indeed, one may assume that master equations are implemented by a finely tuned network of lateral and feed-back interactions. One advantage of this recursive definition in the master equations is that it gives a simple framework for the implementation of association fields. However, this implementation has the consequence of blurring predictions. We explore another route using the {\sc Condensation} algorithm~\citep{Isard98} which surpasses the above approximation. More, it allows us to explore the role of prediction in solving the aperture problem on a more generic level. %
\section{Model \& methods}%
\subsection{Particle filtering}%
\head{\textbf{weighted sampling:} definition / algo / read-out}%
Master equations can be approximated using Sequential Monte Carlo (SMC). This method (also known as ``particle filters'') is a special version of importance sampling, in which PDFs are represented by weighted samples. Here, we represent joint variables of motion as a set of 4-dimensional vectors merging position $\vec{x}= (x, y)$ and velocity $\vec{V}=(u, v)$. Using sampling, any distribution $p(\vec{x}, \vec{V} | I)$ may be approximated by a set of weighted samples, or ``particles'' $\pi^{1:N} = \{ \pi^i \}_{i\in1:N} = \{ (x^{i}, y^{i}, u^{i}, v^{i})\}_{i\in1:N}$ along with weights $w^{1:N} = \{ w^{i} \}_{i\in1:N}$. Weights are positive ($\forall i, w^{i} \geq 0$) and normalized ($ \sum^N_{i=1}w^{i} =1$). By definition, $p(\vec{x}, \vec{V} | I)\approx \hat{p}(\vec{x}, \vec{V} | I)$ with 
\begin{align}
\hat{p}(\vec{x}, \vec{V} | I) & = \sum_{i\in1:N} w^{i} \cdot \delta( \vec{x} - (x^{i}, y^{i}), \vec{V} - (u^{i}, v^{i}) ) \label{eq:sampling}
\end{align}
where $\delta$ is the Dirac measure. There are many different sampling solutions to one given PDF. Prototypical solutions are either a uniform sampling of position and velocity spaces with weights proportional to $p(\vec{x}, \vec{V} | I)$ or the sampling corresponding to uniform weights with a density of samples proportional to the PDF. Compared to other approximations, such as the Laplacian approximation of the PDF by a Gaussian, this representation has the advantage to allow the representation of arbitrary distributions, such as the sparse or multimodal distributions that are often encountered with natural scenes.%

\head{\textbf{CONDENSATION1: } init / drift + diffusion }%
This weighted sample representation makes the implementation of Equations~\eqref{eq:mrf1}-\eqref{eq:mrf2} tractable on a sequential computer. To initialize the algorithm, we set particles $\pi^{1:N}_{t=0}$ to random values with uniform weights. Then, the following two steps are repeated in order to recursively compute particles $\pi^{1:N}_{t}$. This set represents $\hat{p}(\vec{x}_{t}, \vec{V}_{t} | I_{0:t})$ while particles $\pi^{1:N}_{t-dt}$ represent $\hat{p}(\vec{x}_{t-dt}, \vec{V}_{t-dt} | I_{0:t-dt})$. First, the prediction equation implemented by Equation~\eqref{eq:mrf1} and that uses the prior predictive knowledge on the smoothness of the trajectory may be implemented to each particle by a deterministic shift followed by a diffusion such as defined in the generative model (Equations~\eqref{eq:dyn1}-\eqref{eq:dyn2}). The noise $\mathcal{N}_{\pi}=(\mathcal{N}_{\vec{x}},\mathcal{N}_{\vec{V}})$ is here described as a 4-dimensional centered and de-correlated Gaussian. This intermediate set of particles represents an approximation of $p(\vec{x}_{t}, \vec{V}_{t} | I_{0:t-dt})$.

\head{\textbf{CONDENSATION2: } filtering + loop TODO: parler de l'hypothese ergodique: pas besoin de populations de particules }%
Second, we update measures as recorded by the observation likelihood distribution $\hat{p}(I_{t} | \vec{x}_{t}, \vec{V}_{t})$ that is computed from the input sensory flow. As in the SMC algorithm, we apply Equation~\eqref{eq:mrf2} using the sampling approximation 
by updating the weights of the particles: $\forall i$, %
\begin{equation}
 w^{i}_{t} = 1 / Z \cdot w^{i}_{t-dt} \cdot p(I_t | \pi^{i}_{t})
\end{equation} 
where the scalar $Z$ ensures normalization of the weights ($ \sum^N_{i=1}w^{i}_{t} =1$). Likelihood $p(I_{t} | \pi^{i}_{t})$ is computed using Equation~\eqref{eq:obs} thanks to a standard method that we describe below. Last, a usual numerical problem with SMC is the presence of particles with small weights due to sample impoverishment. We use a classical resampling method: This procedure results in eliminating particles that assign little weights (as detected by a threshold relative to the average weight), and duplicating particles that assign largest weights. In analogy with an homeostatic transform, it has the property of distributing resources while not changing the representation~\citep{Perrinet10}. In summary, this whole formulation is similar to the processing steps in the {\sc Condensation} algorithm which was used in another context for the tracking of moving shapes~\citep{Isard98} and we apply it here to motion-based prediction as defined in Equations~\eqref{eq:dyn1}-\eqref{eq:dyn2}. Note that though our probabilistic approach is exactly similar to that of~\citep{Burgi00} at the computational and algorithmic levels, the actual implementation is completely different. The approach of~\citet{Burgi00} seeks to achieve an implementation close to neural networks. We will see that the particle filtering approach, though a priori less neuromorphic,  allows to achieve a higher precision in motion-based prediction which is essential in observing the emergence of complex behaviors characteristics of neural computations. %

An advantage of importance sampling it that it allows to easily compute moments of the distribution. This is particularly useful to define different read-out mechanisms in order to compare the output of our model with biological data. For instance, we can compute the read-out for tracking eye movements as the best estimator (that is the conditional mean) using the approximation of $p(\vec{x}, \vec{V} | I)$:
\begin{align}
<\vec{V} | I> = \int p(\vec{x}, \vec{V} | I) \cdot \vec{V} \cdot d\vec{V} \cdot d\vec{x} \approx \begin{pmatrix} \sum^N_{i=1} w^{i} \cdot u^{i}\\ \sum^N_{i=1} w^{i} \cdot v^{i} \end{pmatrix} \label{eq:readout}
\end{align} 
Furthermore, by restricting the integration to a sub-population of neurons, we can also compare model output with single neuron selectivity and thus test how neuronal properties such as contrast gain control or center-surround interactions could emerge from such predictive coding.%
\subsection{Numerical simulations}%
\head{\textbf{SMC:} N / resampling parameter }%
The SMC algorithm itself is controlled by only two parameters. The first one is the number of particles $N$ which tunes the algorithmic complexity of the representation. In general, $N$ should be big enough and an order of magnitude of $N \approx 2^{10}$ was always sufficient in our simulations.  In the experimental settings that we have defined here (moving dots or lines), the complexity of the scene is controlled and low. Control experiments have tested the behavior for different number of particles (from $2^5$ to $2^{16}$) and have shown, except for $N$ smaller than $100$, that results were always similar. However, we kept $N$ to this quite high value to keep the generality of the results for further extensions of the model. The other parameter is the threshold for which particles are resampled. We found that this parameter had little qualitative influence providing that its value is large enough to avoid staying in a local minima. Typically, a resampling threshold of $20\%$ was sufficient. %

\head{\textbf{Likelihood:} image space / computation / parameters}%
Once the parameters of the SMC were fixed, the only free parameters of the system were the variances used to define the likelihood and the noise model $\mathcal{N}_{\pi}$. Likelihood of sensory motion was computed using Equation~\eqref{eq:obs} using the same method as~\citet{Weiss02}. We defined space and time as the regular grid on the toroidal space to avoid border effects. Next, visual inputs were $128 \times 128$ grayscale images on $256$ frames. All dimensions were set in arbitrary units and we defined speed such that $V=1$ corresponds in toroidal space to the velocity of one spatial period within one temporal period that we defined arbitrarily to \unit[100]{ms} biological time. Raw images were preprocessed (whitening, normalization) and we computed at each processing step the likelihood locally at each point of the particle set. This computation was dependent only upon image contrast and the width of the receptive field over which likelihood was integrated. We tested different parameters values that resulted in different motion direction or spatio-temporal resolution selectivities. For instance, a larger receptive field size gave a better estimate of velocity but a poorer precision for position, and reciprocally. Therefore, we set the receptive fields size to a value yielding to a good trade-off between precision and locality (that is $5\%$ of the image's width in our simulations). Similarly, likelihood's contrast was tuned to match average noise value in the set of images. We also controlled that using a prior favoring slow speeds had little qualitative influence on our results and we used a flat prior on speeds throughout this manuscript. Once fixed, these two values were kept constant across all simulations.  Note that the individual measurements of the likelihood may represent multi-modal densities if the corresponding individual motions are further than an order of the receptive field's size (as when tracking multiple dots). However, such measurements may be perturbed if individual motions are superimposed on a receptive field. Such a generative model of the input may be accounted for by using a Gaussian mixture model~\citep{Isard98}. The types of stimuli we are considering are always well described by an unimodal distribution and we will here restrict ourselves to this simple formulation.%

\head{\textbf{python:} }%
All simulations were performed using python with modules numpy~\citep{Oliphant07} and scipy (respectively version 2.6, 1.5.1 and 0.8.0) on a cluster of linux nodes. Visualisation was performed using matplotlib~\citep{Hunter07}. All scripts are available upon request from the corresponding author.%
\section{Results}%
\subsection{Prediction is sufficient to solve the aperture problem}%
\head{\textbf{slanted line}: definition / role of line length / aperture solved}%
Similarly to classical studies on the biological solution to the aperture problem, we first used as input the image of a horizontally moving diagonal bar. The initial representation shows a bias towards the perpendicular of the line, as previously found with neuronal~\citep{Pack01,Pei10}, behavioral and perceptual responses~\citep{Lorenceau92,Born06,Masson02a} {(see Figure~\ref{fig:one}-A)}. Moreover, the global motion estimation represented by the probability density function converges quickly to the physical motion, both in terms of retinotopic position and velocity {(see Figure~\ref{fig:one}-C)}. Changing the length of the line did not qualitatively change the dynamics but rather proportionally scaled the time it takes to the system for converging to the physical solution~\citep{Castet93a,Born06} (see Figure~\ref{fig:one}-D). This result demonstrates that motion-based prediction is sufficient to resolve the aperture problem. %

\head{\textbf{observations}: 2D / 2D motion system? / tuned to predictability TODO: really make Julian's experiment! }%
Interestingly, results show that line endings are preferentially tracked (as will be described in Section~\ref{sec:2D}). In fact, the system responds optimally to predictable features and thus, it can correctly detect line endings motion with a probability that is higher than observed for any points located at, say, the middle of the line segment. Moreover, it was shown behaviorally that when blurring the stimulus' line endings, motion representation still converges towards the physical motion albeit with a slower dynamics~\citep{Wallace05}. This is another key signature that we successfully replicated in our model. This shows that end-stopped cells (or more generally local 2D motion detectors~\citep{Wilson92}) are not necessary to solve the aperture problem. On the contrary, a reliable 2D tracking motion system appears to be rather no more than the consequence of cells tuned to predictive trajectories.%

This result has some generic consequences that were not described in previous models such as~\citep{Burgi00,Bayerl07}. First the emergence of line-ending detectors is caused by the fact that the model filters coherent motion trajectories. This property emerges as line-endings follow a coherent trajectory, but this property is therefore not limited to line-endings. As a consequence, the most salient difference is that ``interesting'' features are defined not by a property of the luminance profile but rather by the coherence of their motion's trajectory. Such a distinction is important with regard to biological experiments. Indeed, at the behavioral level,~\citet{Watamaniuk95} have shown that the sequential detection of line-endings is not sufficient to explain at the global level the change of behavior when an object moves on a coherent trajectory. Second, at the physiological level,~\citet{Pack03} have shown in the macaque monkey different phases in the dynamics of MT neurons tuned for line-endings.This suggests that the selective response to line-endings is a consequence of the presentation of a coherent trajectory. %
\begin{figure}%
\centering{\includegraphics[width=.99\columnwidth]{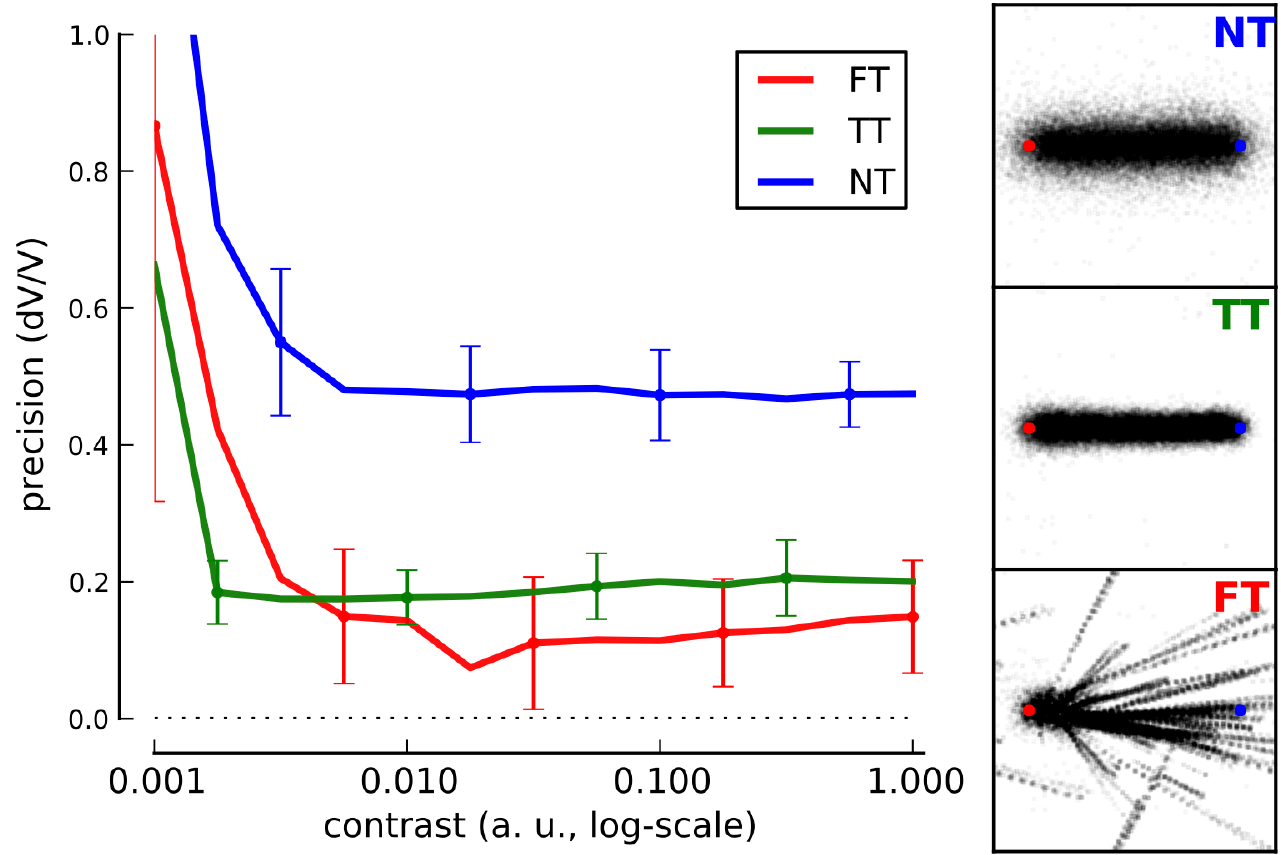}}
\caption{To explore the state-space of the dynamical system, we simulated motion-based prediction for a simple small dot (size 2.5\% of a spatial period) moving horizontally from the left to the right of the screen. We tested different levels of sensory noise with respect to different levels of internal noise, that is, to different values of the strength of prediction. \textbf{(Right)}  Results show the emergence of different states for different prediction precisions: a regime when prediction is weak and which shows high tracking error and variability (No Tracking - NT), a phase for intermediate values of prediction strength (as in Figure~\ref{fig:one}) exhibiting a low tracking error and low variability in the tracking phase (True Tracking - TT) and finally a phase corresponding to higher precisions with relatively efficient mean detection but high variability (False Tracking - FT). We give 3 representative examples of the emerging states at one contrast level ($C=0.1$) with starting (red) and ending (blue) points and respectively NT, TT and FT by showing inferred trajectories for each trial. \textbf{(Left)} We define tracking error as the ratio between detected speed and target speed and we plot it with respect to the stimulus contrast as given by the inverse of sensory noise. Error bars give the variability in tracking error as averaged over 20 trials. As prediction strength increases, there is a transition from smooth contrast response function (NT) to more binary responses (TT and FT). %
\label{fig:three}
}%
\end{figure}%
\subsection{Emergence of texture-independent motion trackers}
\head{\textbf{using dots}: definition / motion-based motion tracker / definition of tracking behavior} %
To further understand these mechanisms, we tested the response of the dynamical system to a coherently moving dot. This was defined as a Gaussian blob of luminance. Its center moved with a constant translational velocity. For a wide range of parameters, we found that the particles representing the distribution of motion quickly concentrate on the dot's centre while their velocity converged to the true physical velocity. Thanks to the additional information given by the predictive information, this convergence is much quicker than what would be obtained by simply  integrating temporally the raw inputs. Moreover, the response of the system is qualitatively different from what is expected in absence of prediction. In fact, if the dot's motion is coherent with the predictive generative model, information is either amplified or reduced, resulting in a progressively more and more binary response as time progresses. This behavior is the consequence of the auto-referential formulation of our motion detection scheme. Indeed, precision of motion estimation is modulated by a prediction that is itself estimated using motion. We therefore see the emergence of a basic tracking behavior where the dot's trajectory is ``captured'' by the system. %

\head{\textbf{tracking threshold}: changing \textit{(1)} noise \textit{(2)} prediction precision / statespace}%
We explored the effects of some key parameters on the tracking behavior of the model. First, when progressively adding uniform Gaussian white noise to the stimulus, we found that convergence time to veridical tracking increased with respect to the level of noise. Then, at a certain level of noise, error bias in the prediction becomes larger than required for the balance in tracking amplification and therefore dots are rapidly lost. This can define a ``tracking sensitivity threshold'' that can be characterized by plotting the contrast response function of our system (see Figure~\ref{fig:three}). Second, we varied the precision of prediction. It is quantitatively defined by the inverse variance of the noise present in the generative model (Equation~\ref{eq:dyn1}-\ref{eq:dyn2}). We observed that convergence speed of tracking grew proportionally with this parameter. For very low precision values,  the tracking behavior is lost. Moreover, we observed that increasing this prediction's precision above a certain threshold leads to the detection of false positives: An initial movement may be predicted in a false trajectory but is not discarded by sensory data. In fact, this is due to the high positive feedback generated by the high precision assigned to the prediction. In summary, varying both parameters, that is, external and internal variability, we can identify three distinct regimes in this state-space: an area of correct tracking (see Figure~\ref{fig:three}-TT), an area where there is no tracking due to low precision or high noise (see Figure~\ref{fig:three}-NT), and an area of false tracking (see Figure~\ref{fig:three}-FT). These three regimes fully characterize the emergence of the tracking behavior of the dynamical system implementing motion-based prediction. %

\head{\textbf{texture independence} spatial textures/ second-order  / tracking of motion - }%
We then studied how such a tracking behavior is independent from the luminance profile of the object being tracked. To achieve that, we tested our system with the same dot but whose envelope was multiplied by a random white noise texture. When this texture consists of a static grating, we obtain one instance of second-order motion (see~\citep{Lu01} for a review). Although the convergence was longer and more variable, tracking was still observed in a robust fashion and the envelope's motion was ultimately retrieved. This property is due to the fact that, in the generative model, we define the prediction as based on both motion's position and trajectory, independently of the local geometry of image features. This is different from motion detection models which rather try to track a particular luminance feature~\citep{Wilson92,Lu01}. As a consequence, this dynamical system will have a preference for objects conserving their motion along a trajectory, independently of their texture. Such invariance is usually obtained by introducing, and tuning, a well-known static non-linear computation such as divisive normalization~\citep{Simoncelli98,Rust06}%
\subsection{Role of context for solving the aperture problem}
\label{sec:2D}%
\head{\textbf{object interference}: diagonal line / gain-control / explaining away}
In order to better understand how the different parts of the line interact in time, we finally investigated modulation of neighboring motions in the aperture problem. In fact, this also corresponds to the case of the diagonal line: At the initial time step, motion position information is spread preferentially along the edge of the line and represents motion ambiguity with a speed probability distributed along the constraint line (see Figure~\ref{fig:one}-A, Inset). In particular, trajectories are inferred on different trajectories preferentially on the points of the line but with directions which are initially ambiguous due to the aperture problem.  These different trajectories evolve independently but are ultimately in competition. To understand the underlying mechanism, we first focus on a single step of the algorithm at three independent key positions of the stimulus: the two edges and the center. Compared with the case without prediction, we show that prediction induces a contextual modulation of the response to different trajectories, such as explaining away trajectories that fall off the line (see Figure~\ref{fig:four}-Top). This modulation acts on a large scale as a gain-control mechanism which is reminiscent to what is observed in center-surround stimulation. 

\head{\textbf{diffusion} line / explain by looking @ backwards propagation / aperture}
We can then analyze in greater details the dynamics of motion distributions for the aperture problem. From the initial step, unambiguous line-ending information spreads progressively towards the rest of the line, progressively explaining away motion signals that are inconsistent with the target speed (see Figure~\ref{fig:four}-Bottom). In fact, from the formulation of prediction in the master equation, probability at a given point reflects the accumulated evidence of each trajectory leading to that point as it is computed by the predictive prior. Combined with likelihood measurements, incoherent trajectories will be progressively explained away as they fall off the line. Such gradual diffusion of information between nearby locations explains the role of line length already documented in Figure~\ref{fig:one}-D, as well as why information takes time to diffuse at the global scale of the stimulus, as is reported at the physiological level,~\citet{Pack03}. In summary, contrary to other models consisting of a selection stage, the system selects coherent features in an autonomous and progressive manner based on the coherence of all their possible trajectories. This ultimately explains why in the aperture problem, information diffuses in the system from line endings to the rest of the segment to ultimately resolve the correct physical motion. %

A counter-intuitive result is that the leading bottom line-ending is less informative that the trailing upper line-ending. This was already evident from the asymmetry revealed in Figure~\ref{fig:four}-Top which explicits that motion-based prediction will have a different effect on both line-endings. Indeed, in the leading line-ending, most information is diffused to the rest of the line and is not explained away. On the contrary, for the trailing line-ending, the diffusion of information is more constrained as any motion hypothesized to be going upwards would soon be explained away from motion-based prediction as it would fall off the line. This asymmetry is clearly observable in  Figure~\ref{fig:four}-Bottom as the ambiguous information (coded here by a blueish hue) is progressively resolved by the diffusion of the information originating from the trailing line-ending. Unfortunately, the experiments using blurring of the line performed by~\citet{Wallace05} were preformed symmetrically, that is, similarly for both edges. We thus predict that blurring the trailing line-ending only should lead to a greater bias angle as blurring the leading line-ending only. %

\begin{FPfigure}%
\centering{\includegraphics[width=\textwidth]{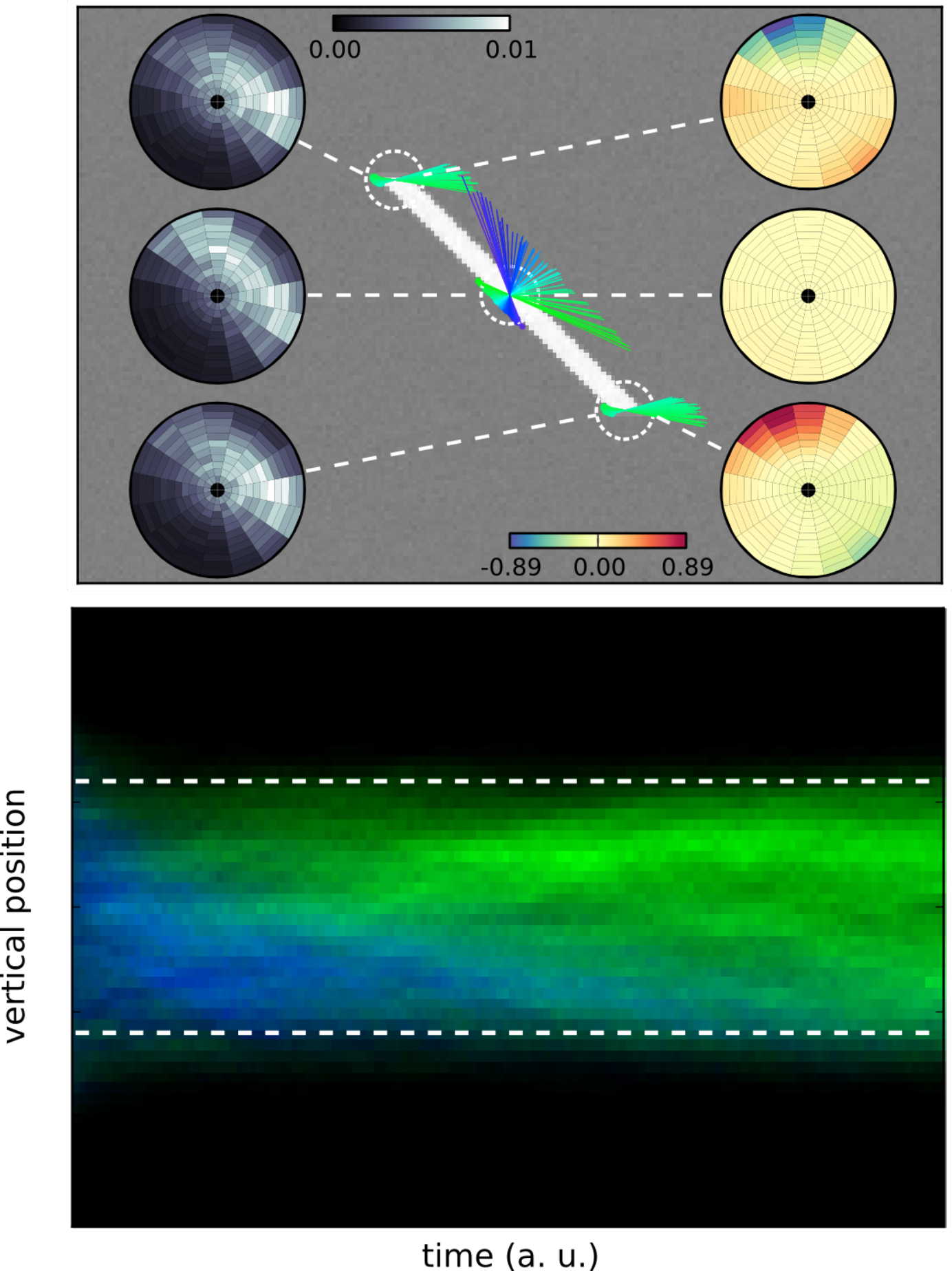}}%
\caption{%
\textbf{(Top)} Prediction implements a competition between different trajectories. Here, we focus on one step of the algorithm by testing different trajectories at three key positions of the segment stimulus: the two edges and the center (dashed circles). Compared to the pure sensory velocity likelihood (left insets in grayscale), prediction modulates response as shown by the velocity vectors (direction coded as hue as in Figure~\ref{fig:one}) and by the ratio of velocity probabilities (log ratio in bits, right insets). There is no change for the middle of the segment (yellow tone), but trajectories that are predicted out of the line are ``explained away'' (navy tone) while others may be amplified (orange tone). Notice the asymmetry between both edges, the upper edge carrying a suppressive predictive information while the bottom edge diffuses coherent motion. \textbf{(Bottom)} Finally, the aperture problem is solved due to the repeated application of this spatio-temporal contextual information modulation. To highlight the anisotropic diffusion of information over the rest of the line, we plot as a function of time (horizontal axis) the histogram of the detected motion marginalized over horizontal positions (vertical axis), while detected direction of velocity is given by the distribution of hues. Blueish colors correspond to the direction perpendicular to the diagonal while a green color represents a disambiguated motion to the right (as in Figure~\ref{fig:one}). The plot shows that motion is disambiguated by progressively explaining away incoherent motion. Note the asymmetry in the propagation of coherent information. 
\label{fig:four}%
}%
\end{FPfigure}%
\section{Discussion}%
\head{\textbf{summary :} aperture problem / prediction are important / we are first one to show emergence of visual computation}%
Our computational model shows that motion-based prediction is sufficient to solve the aperture problem as well as other motion integration phenomena. The aperture problem instantiated with slanted lines in visual space helps to capture several generic computations which are often considered as essential features of any sensory areas. We have shown that predictive coding through diffusion is sufficient to explain the emergence of local 2D motion detectors but also texture-independent motion grabbers. It can also implement context-dependent competition between local motion signals. All these computations are emerging properties from the dynamics of the system. This view is opposite to the classical assumptions that these mechanisms are implemented by specific, separated mechanisms (e.g.~\citep{Wilson92,Lu01,Grossberg01,Tsui10}). Instead, we demonstrate herein that all these properties must be seen as the mere consequence of a simple, unifying computational principle. By implementing a predictive field, motion information is anisotropically propagated as modulated by sensory, local estimations such that motion representation dynamically diffuses from a local to a global scale. This model offers a simplification of our original model proposed earlier~\citep{Tlapale10vr}. %
 
\subsection{Relation to other models}%
In fact, we can take advantage of the work from~\citet{Tlapale10benchmark} to compare our model with~\citep{Tlapale10vr} and a large range of models in the community. This study compared the results obtained from different modeling approaches on the same aperture problem and used their model as a reference point. Taking this study as a reference, there are two main difference with our model. First, it does not try to make a neuromorphic approach except the fact that (to respect the definition of the aperture problem) information is grabbed locally and propagated on a neighborhood. Moreover, in our model, information is represented explicitly by probabilities and we make no assumption on how it is represented in the neural activity as this would introduce unnecessary hypothesis regarding our objective. Second, motion-based prediction defines an anisotropic, context-dependent direction of propagation while most previous models were using an isotropic diffusion dependent on some feature characteristics (like gating the diffusion by luminance). However, our model uses explicitly the selectivity brought by the anisotropic diffusion. As a consequence it needs less tuning of the parameters of the diffusion mechanisms, which is a common problem in the latter type of models. A further advantage of our approach is that it does not contradict previous models. Rather, motion-based prediction seem to be a promising approach to be implemented in neuromorphic models. 

\head{\textbf{Relation to other models:} similar in behavior / it's a framework for future models gestalt laws}%
In particular, several parts of our model are similar to previous models of motion detection but its whole implementation is radically novel. %
First, it inherits from properties of functional models such as the probabilistic formulation of~\citet{Weiss02} but with more simple hypotheses. For instance, we do not need a prior distribution favoring slow speeds or some selective process that are needed to pre-process the data~\citep{Weiss02,Barthelemy08}. Our model uses a simple Markov Chain formulation which has been used for spatial luminance-based prediction or shape tracking with SMC in the {\sc Condensation} algorithm~\citep{Isard98}, but this was to our knowledge not applied to an explicit definition of motion-based prediction. Note that the model presented in~\citep{Bayerl07} includes an anisotropic diffusion based on motion-based prediction but that this study was using a neural approximation of the kind of~\citep{Burgi00}. However, they did not study in particular the role of prediction in the progressive resolution of the aperture problem and its characteristic signature compared to biological data. The application of their fast implementation to our model appears to be a promising perspective. Ultimately, our model also gives a more formal description of the dynamical Bayesian model that we have originally suggested to implement dynamical inference solution for motion integration~\citep{Montagnini07,Bogadhi11}. %

Moreover, when compared to other models designed for understanding visual motion detection~\citep{Wilson92,Grossberg01,Bayerl04}, our approach is more parsimonious as we don't need to explicitly model specialized edge detectors. On the contrary, we show that these local feature detectors must be rather seen as emerging properties from a subset of coherent-motion detectors. Nevertheless, this emergence needs a fine scale prediction as we have shown that these properties depend on prediction's precision. Our computational implementation using SMC could reach higher precision levels compared to the earlier predictive model proposed by~\citet{Burgi00}. We could therefore explore a range of parameters and stimuli (such as the aperture problem) that is radically different from the original study. Moreover, some non-linear behaviors observed in our model are similar to other signatures of linear/non-linear models such as the cascade model from~\citet{Rust06} or mesoscopic models~\citep{Bayerl04,Bayerl07,Tlapale10vr}. However, these last models are specifically tuned by assembling complex and precise knowledge from the dynamical behavior of neurons and their interactions to fit the results that were obtained neurophysiologically. In our model, though, these properties emerge from the interactions in the probabilistic model. %

\subsection{Toward a neural implementation}
More generally, this probabilistic and dynamical approach unveils how complex neural mechanisms observed at population levels (or from their read-outs) may be explained by the interactions between local dynamical rules. As mentioned above, both visual~\citep{Pack01,Pack03,Pack04,Smith10} and somatosensory~\citep{Pei10} systems exhibit similar neuronal dynamics when solving the aperture problem or other sensory integration tasks in space and time. This suggests that different sensory cortices might use similar computational principles for integrating sensory inflow into a coherent, non-ambiguous representation of objects motion. By avoiding specific mechanisms such as neuronal selectivities for some specific local features, our approach offers a more generic framework. It also allows to seek for simple, low-level mechanisms underlying complex visual behavior and their dynamics as observed, for instance with reflexive tracking eye movements (see~\citep{Masson12} for a review). Lastly, we propose that distributions of neural activity on cortical maps act as probabilistic representations of motion over the whole sensory space. This suggests that, for instance in cortical areas V1 and MT, all probable solutions are initially superposed. This is coherent with the dynamics of the population of MT neurons when solving the aperture problem or computing plaid pattern motion~\citep{Pack01,Pack04,Smith10}. Simple decision rules can be applied to these maps to trigger different behaviors such as saccadic and smooth pursuit eye movements as well as perceptual judgements of motion such as direction and speed. Then, the temporal dynamics of these behavioral responses can be explained by the dynamics of predictive coding at sensory stage~\citep{Bogadhi11}.

\head{\textbf{new computational paradigms:} association field /difference with image processing / novel non-von-Neumann approaches / good for neuroscience}%
This work provides new insights for neuroscience but also for novel computational paradigms. %
In fact, biological vision still outperforms any artificial system for simple tasks such as motion segmentation. Our simple model is validated based on neurophysiological and behavioral data and gives several perspectives for its application to image processing. In the future, our model will provide interesting perspectives for exploring novel probabilistic and contextual interactions thanks to the use of neuromorphic implementations. Indeed, it is impossible in practice to implement today the full system on classical von-Neumann architectures due to the size of the memory that is required to implement such complex association fields. However, as we saw above, the probabilistic representation of motion has a natural representation in a neural architecture, where many simple processors are densely connected. Thus, this model is structurally compatible with generic neural architectures and it is a candidate functional implementation on wafer-like hardware. Such recent innovative computing architectures enable to construct specialized neuromorphic systems, allowing new possibilities thanks to their massive parallelism~\citep{Bruderle11}. %
In return, this approach will allow us to implement models simulating complex association fields. Studying novel computational paradigms in such systems will help extend our understanding of neural computations.%
\subsection*{Acknowledgments}
\Acknowledgments

\end{document}